\documentclass[aps,prl,twocolumn,groupedaddress]{revtex4}
\usepackage{graphicx,amsmath}
\begin{document}

\title{Observation of non-local dielectric relaxation in glycerol}
\author{A. A. Pronin$^1$}
\author{K. Trachenko$^{2,3}$}
\author{M. V. Kondrin$^4$}
\author{A. G. Lyapin$^4$}
\author{V. V. Brazhkin$^4$}

\address{$^1$ General Physics Institute, Russian Academy of Sciences, Vavilov Street 38, Moscow 119991, Russian Federation}
\address{$^2$ Department of Earth Sciences, University of Cambridge, Cambridge CB2 3EQ, UK}
\address{$^3$ South East Physics Network and School of Physics, Queen Mary University of London, Mile End Road, London, E1 4NS, UK}
\address{$^4$ Institute for High Pressure Physics, Russian Academy of Sciences, Troitsk, Moscow Region 142190, Russian Federation}

\begin{abstract}
{Since its introduction, liquid viscosity and relaxation time $\tau$ have been considered to be an intrinsic property of the system that is essentially local in nature and therefore independent of system size. We perform dielectric relaxation experiments in glycerol, and find that this is the case at high temperature only. At low temperature, $\tau$ increases with system size and becomes non-local. We discuss the origin of this effect in a picture based on liquid elasticity length, the length over which local relaxation events in a liquid interact via induced elastic waves, and find good agreement between experiment and theory.}
\end{abstract}


\maketitle

Viscosity, $\eta$, is the basic property of matter that quantifies its resistance to flow, and was defined by Newton as the proportionality coefficient between shear stress and velocity gradient. It is commonly used to describe properties of gases, liquids and even solids where flow is often referred to as ``creep'' \cite{orowan}. Viscosity of liquids perhaps stands out in that list, in that theory of liquids in general is much less developed as compared to other phases of matter such as solids or gases. For example, it is stated \cite{landau} that unlike for solids and gases, the most basic property of a liquid such as energy can not be calculated in general form. The stated reason is that interactions in a liquid are strong and system-specific so that the energy strongly depends on system type. At the same time, atomic displacements in a liquid are large. Consequently, neither weakness of interactions as in a gas nor smallness of atomic displacements as in a solid is a small parameter that can be used to obtain general results \cite{landau}.

Similarly, temperature dependence of equally basic liquid property, viscosity, has remained another major theoretical hurdle in condensed matter physics. In particular, the behaviour of $\eta$ at low temperature is at the heart of the problem of glass transition \cite{dyre,angell1,langer,ngai}.

A property closely related to viscosity is liquid relaxation time, $\tau$. It was phenomenologically introduced by Maxwell in the viscoelastic picture of liquid flow as $\eta=G_{\infty}\tau$, where $G_{\infty}$ is instantaneous shear modulus \cite{maxwell}. Frenkel subsequently identified $\tau$ with the average time between two consecutive atomic jumps in a liquid at one point in space \cite{frenkel}. Larger $\tau$ correspond to less frequent atomic jumps and larger viscosities.

Notably, in both earlier \cite{maxwell,frenkel,eyring} and modern \cite{landau-el,zwanzig,hansen,dyre,angell1,landau-hy} discussions, it was assumed that the flow property of a liquid, $\eta$ or $\tau$, is essentially an intrinsic property governed only by the liquid structure and interatomic interactions, and therefore is necessarily local. Applied to viscosity, the assertion of locality implies that as long as there is enough bulk liquid between two moving plates so that surface effects can be ignored, viscosity does not depend on system size. Equally, $\tau$ has been considered to be local and independent of system size \cite{maxwell,frenkel,eyring,hansen,dyre,angell1,landau-hy}. Consequently, several well-known relationships for $\eta$ or $\tau$ such as Stokes law, Einstein-Stokes relationship between viscosity and diffusion and so on are derived and discussed in the picture where $\eta$ and $\tau$ are intrinsic and local \cite{frenkel,eyring,hansen,landau-hy}. The same applies to general Navier-Stokes equations that are used to derive and discuss a wide range of results and effects in liquids \cite{landau-hy}.

In this paper, we perform dielectric relaxation experiments in glycerol and find that $\tau$ is independent on system size at high temperature only, but increases with system size at low temperature, i.e. becomes essentially non-local. We attribute this effect to non-local interactions between atomic jumps in a liquid via induced high-frequency elastic waves. We find both qualitative and quantitative agreement between this theory and present experiments.

We first discuss the theoretical rationale for our experiment. We have recently discussed \cite{jpcm,ser} the origin of several anomalous and intriguing relaxation laws in liquids that are at the centre of the problem of glass transition, such as Vogel-Fulcher-Tammann (VFT) law, stretched-exponential relaxation and dynamic crossovers. An important outcome of this discussion is that relaxation in a liquid necessarily becomes non-local due to interaction between atomic jumps. The non-locality originates as follows.

An elementary flow process in a liquid is the jump of an atom from its surrounding atomic ``cage'', accompanied by large-scale rearrangement of the cage atoms. We call this process a local relaxation event (LRE). A LRE lasts on the order of elementary (Debye) vibration period $\tau_0=0.1$ ps. On this short timescale (as well as any other timescale that is shorter than $\tau$), the surrounding liquid can be viewed as an elastic medium \cite{frenkel}. Hence, large atomic motion from a LRE elastically deforms the surrounding liquid, inducing elastic waves. The wave frequency, $\omega$, is on the order of Debye frequency because the wavelength is on the order of interatomic separations. Therefore, $\omega>\frac{1}{\tau}$ as long as $\tau$ exceeds $\tau_0$, i.e. in the whole range of $\tau$ up to $\tau=10^2-10^3$ s at the glass transition temperature. As discussed by Frenkel \cite{frenkel}, waves with frequency $\omega>\frac{1}{\tau}$ are propagating in a liquid as in a solid. The waves distort cages around other LRE centres in the liquid, and therefore affect their relaxation.

Hence, we identified LRE-induced elastic waves as the physical mechanism of mediating interactions in a liquid, and proposed \cite{jpcm} that this interaction sets the cooperativity of relaxation whose origin was widely discussed but not understood from the physical point of view \cite{dyre,angell1,langer,ngai}.

The key question is the range of this interaction, or the propagation range of high-frequency elastic waves. As discussed in detail \cite{jpcm}, this range is given by $d_{\rm el}$:

\begin{equation}
d_{\rm el}=c\tau
\label{del}
\end{equation}

\noindent where $c$ is the speed of sound.

The non-trivial point is that $d_{\rm el}=c\tau$ {\it increases} with $\tau$. This is directly opposite to the commonly discussed decay of hydrodynamic waves, whose propagation range varies as $1/\tau$. The difference is due to the solid-like regime of wave propagation, $\omega\tau>1$, which is qualitatively different from the hydrodynamic regime, $\omega\tau<1$ \cite{frenkel} (see Ref. \cite{jpcm} for a detailed discussion). We called $d_{\rm el}$ liquid elasticity length because it defines the range over which two LREs interact with each other via induced elastic waves. Importantly, $d_{\rm el}=c\tau$ {\it increases} on lowering the temperature because $\tau$ increases. We proposed that this is the key to the emergence of slow relaxation, VFT law and dynamic crossovers \cite{jpcm,ser}. In this picture, the increase of $d_{\rm el}$ and concomitant cooperativity of relaxation increases the activation barriers for LREs and $\tau$ \cite{jpcm}.

We note that $d_{\rm el}$ is different from the solidity length, $l$, introduced in Ref. \cite{dyre1}. In particular, $l$ was derived not as the propagation range of high-frequency waves in a liquid, but as a distance that the sound wave travels without LREs appearing anywhere along its path during time $\frac{l}{c}$. As a result, $l$ is proportional to $\tau^{\frac{1}{4}}$ (compare with Eq. (\ref{del}) where $d_{\rm el}\propto\tau$), is significantly smaller than $d_{\rm el}$, and is limited by hundreds of nm at $T_g$ \cite{dyre1}.

As long as $d_{\rm el}<L$, where $L$ is system size, relaxation effects are not affected by system size. On the other hand, when $d_{\rm el}\ge L$ at low temperature, one expects to find that relaxation depends on system size, provided $d_{\rm el}<d_t$. Here, $d_t$, phonon thermalization length, is the distance (typically up to several mm at room temperature \cite{d1,d2,d3}) over which an induced phonon thermalizes due to anharmonicity, with the result that the phonon spectrum of the system becomes the same as in thermal equilibrium. In particular, two important system size effects are immediately predicted by this theory. First, the theory predicts that when $d_{\rm el}=L$, $\tau$ crosses over from the VFT law to a more Arrhenius dependence, because at this point further increase of $d_{\rm el}$ does not increase the number of interacting LREs in the system \cite{jpcm}. This effect has been observed experimentally in a large number of glass-forming liquids \cite{roland,sti,sti1}.

Second, this picture predicts that when $d_{\rm el}\ge L$, $\tau$ increases with $L$ and becomes essentially non-local. We note that in the above discussion, $\tau$ can be considered non-local when $d_{\rm el}<L$ as well, as long as $d_{\rm el}$ exceeds the distance between neighbouring LREs. However, experimental verification of the non-locality of $\tau$ is most easily verified by varying $L$ and demonstrating that $\tau$ increases with $L$. The first demonstration of this effect in a macroscopic system was done in the series of Stokes viscosity experiments \cite{honey}. Recently, a similar non-locality of viscosity was found in molecular dynamics simulations \cite{leva}. We now discuss the results of our dielectric relaxation experiment supporting the prediction of the size effect.

We have measured $\tau$ in glycerol, a commonly studied glass-forming liquid, using well-known broadband (10-10$^8$ Hz in the current experiments) dielectric spectroscopy technique in capacitors of two different sizes. Several important experimental conditions need to be met in order to study size effect. These include ensuring that temperature difference between the two capacitors is minimized, as well as temperature gradients in each system. Next, the liquid in each capacitor should be a closed system, and not connected to a liquid of larger size, as can be the case in dielectric relaxation experiments. Further, pressure variations in a closed system related to thermal expansion should be avoided because they affect $\tau$.

To meet these requirements, we have designed and built the test cell shown in the inset in Figure 1. The cell contained two switchable capacitors C1 and C2, both of the same radius $R=0.75$ mm and different heights $h_1=0.2$ mm and $h_2=0.02$ mm, respectively. The distance between the capacitor plates is maintained by flat teflon spacers. Both capacitors were cooled simultaneously. Subsequently, their temperature was stabilized by a temperature controller with accuracy better than 0.02 K. Due to efficient thermal coupling between the capacitors, temperature gradient through the test cell was not larger than 0.01 K.

Samples of ultra-pure ($>$99\%) glycerol were purchased from ICN Biomedicals. High sample purity level was additionally checked by refractometry. At each temperature, the liquid was equilibrated during temperature stabilization lasting several minutes, considerably exceeding stress relaxation time in our temperature range. Frequency dependencies of the impedance of the capacitors $Z$($F$) were measured by QuadTech 7600 precision LCR Meter below 1 MHz and by Hewlett-Packard HP4191A reflectometer above 1 MHz. Typical dielectric loss spectra at the $T=$233 K, where the maximum value of $F_m=\frac{1}{\tau}$ is in the kHz range, is shown in Figure 1.

\begin{figure}
\begin{center}
{\scalebox{0.65}{\includegraphics{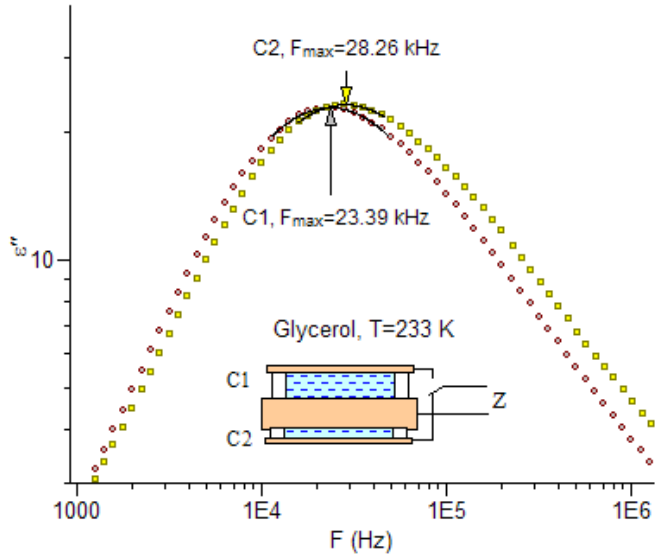}}}
\end{center}
\caption{Schematic experimental setup (inset) and a typical dielectric loss spectra of glycerol measured in the larger (circles) and smaller (squares) capacitor at T=233 K.}
\end{figure}

\begin{figure}
\begin{center}
{\scalebox{0.65}{\includegraphics{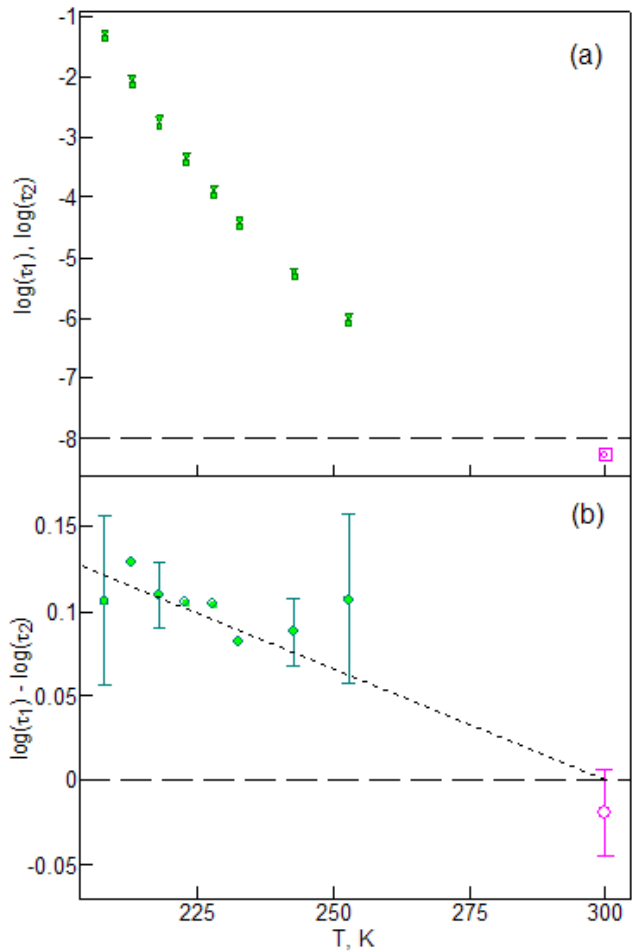}}}
\end{center}
\caption{(a) $\tau$ in large (upper symbols) and small (lower symbols) capacitors. The horizontal long-dashed line corresponds to $\tau$ above which the size effect is predicted to exist. (b) $\log(\tau_1)-\log(\tau_2)$, where $\tau_1$ and $\tau_2$ are relaxation times in large and small capacitors, respectively.
}
\end{figure}

In Figure 2, we plot $\tau$ of both capacitors as a function of temperature, together with $\log(\tau_1)-\log(\tau_2)$, where $\tau_1$ and $\tau_2$ are relaxation times in large and small capacitors, respectively. Figure 2 makes three important points about the predicted size effect.

First, the size effect is absent at high temperature, 300 K. According to Figure 2a, this temperature corresponds to $\tau\lesssim 10^{-8}$ s and hence $d_{\rm el}\lesssim$ 0.02 mm ($c\approx$2000 m/s for glycerol). Therefore, $d_{\rm el}$ is just smaller than the smallest dimension of both capacitors ($h_2=0.02$ mm). According to the theoretical prediction, $\tau$ should be the same in both systems: $\tau_1=\tau_2$, in agreement with Figure 2b.

Second, we observe that the size effect is present at low temperature, below 250 K. At 250 K, $\tau\approx 10^{-6}$ s (see Figure 2a), giving $d_{\rm el}\approx$ 2 mm. Therefore, $d_{\rm el}$ is larger than the largest dimension of both capacitors below 250 K. According to theoretical prediction, $\tau$ should be larger in the larger system in this temperature range: $\tau_1>\tau_2$. This is in agreement with the experimental results in Figure 2b.

Third and finally, we calculate the magnitude of the size effect from the theory. According to the theory \cite{jpcm}, the activation barrier for a LRE, $U$, is

\begin{equation}
U=U_0+\frac{T_0}{4\pi}\int\frac{\rm{d}V}{r^3}
\label{u1}
\end{equation}

The first term in Eq. (\ref{u1}), $U_0$, is the high-temperature intrinsic, or non-cooperative, activation barrier for a LRE than depends on structure and the type of interactions in a liquid but not on the interaction with other LREs. The second term describes the effect of elastic interaction between LREs.  Here, $T_0$ is the VFT temperature, and $\frac{1}{r^3}$ represents the decay of elastic stress. The integration is performed over the volume inside the system boundaries if $d_{\rm el}\geq L$ and inside $d_{\rm el}$ if $d_{\rm el}<L$. We note that if $d_{\rm el}<L$, $U$ is temperature-dependent because it depends on $d_{\rm el}$ and therefore on $\tau$, giving the VFT law for $\tau$ \cite{jpcm}. The lower integration limit in Eq. (\ref{u1}) is given by the size of a local rearranging region, $d_0$ ($d_0\approx 10$ \AA).

Integrating in cylindrical coordinates gives

\begin{equation}
U=U_0+\frac{T_0}{2}\ln\frac{2hR}{d_0(h+\sqrt{h^2+R^2})}
\label{u2}
\end{equation}

\noindent where we have taken into account that $d_0\ll h_1, h_2, R$.

We define the magnitude of the size effect, $\Delta$, as $\Delta=\frac{\tau_1}{\tau_2}$, where $\tau_1$ and $\tau_2$ are relaxation times in larger and smaller capacitors, respectively. Using Eq. (\ref{u2}) and $\ln(\tau)=\ln(\tau_0)+\frac{U}{T}$, where $\tau_0$ is Debye vibration period, we write:

\begin{equation}
\ln\Delta=\frac{T_0}{2T}\ln\left(\frac{h_1}{h_2}\frac{h_2+\sqrt{h_2^2+R^2}}{h_1+\sqrt{h_1^2+R^2}}\right)
\label{eff}
\end{equation}

\noindent

In the low temperature range where the size effect is seen in Figure 2b (208--253 K), Eq. (\ref{eff}) gives $\Delta=1.7-1.9$, where we have used $T_0=135$ K \cite{t0} and the above values of $h_1$, $h_2$ and $R$. The experimental values of $\Delta$ from Figure 2b are in the range 1.2--1.3. Therefore, the theoretical prediction of the size effect is in reasonable agreement with the experimental value, providing, importantly, a correct order of magnitude of the effect. We observe that $\tau_1$ and $\tau_2$ themselves vary by many orders of magnitude in the above temperature range (see Figure 2a), hence a reasonably good agreement of $\frac{\tau_1}{\tau_2}$ between theory and experiment is particularly encouraging.

We note that Eq. (\ref{eff}) predicts the decrease of $\Delta$ with temperature. Consistent with this, there is a decreasing trend of $\Delta$ in Figure 2b despite the larger error bars at the boundaries of the low temperature range, because the errors in the middle of the range were smaller in our experiments.

We attribute the higher theoretical value of $\Delta$ to the decay of LRE-induced waves due to anharmonicity \cite{d1,d2,d3}. This decay is not considered in the theory, and leads to the reduction of the wave amplitude within $d_{\rm el}$ and therefore weakening of the size effect.

We finally note that $T_g$ decreases, compared to the bulk values, in liquids confined to nano-scale pores as well as in thin polymer films, implying that $\tau$ becomes smaller in these systems \cite{mckenna}. This effect is therefore consistent with our theory. However, as reviewed in Ref. \cite{mckenna} in detail, it has remained unclear whether this is an intrinsic effect and to what extent the observed size effects are related to extrinsic factors such as the interaction between the liquid and the surface, particularly important in small systems. Our results, on the other hand, are done on essentially macroscopic samples where the above extrinsic effects are expected to be small. Our results therefore support the existence of a genuine intrinsic size effect in liquids, in accordance with our recent theory.

In summary, our dielectric relaxation experiments show that relaxation time of glycerol increases with system size at low temperature, implying that $\tau$ (and similarly, viscosity) become non-local. We propose that the origin of this unexpected behaviour is the long-range elastic interaction between local relaxation events, whose range increases on lowering the temperature and extends to system size. The results of our experiments are in a good qualitative and quantitative agreement with theoretical predictions.

We are grateful to R. Casalini, C. M. Roland, M. T. Dove and V. Heine for discussions and to EPSRC for support.

\end{document}